\documentclass[a4paper,fleqn,usenatbib]{mnras}
\usepackage[T1]{fontenc}
\usepackage{ae,aecompl}
\usepackage{amsmath}
\usepackage{amsfonts}
\usepackage{amssymb}
\usepackage{graphicx}
\usepackage{color,xcolor}
\def\mnras{MNRAS}
\def\aap{A$\&$A}
\def\nat{Nature}
\def\apj{ApJ}
\def\apjl{ApJ Letters}
\def\apjs{ApJS}

\def\aj{AJ}
\def\icarus{Icarus}

\newcommand{\me}{\, {\rm M}_{\oplus}}
\newcommand{\msun}{\, {\rm M}_{\odot}}

\newcommand{\au}{\, {\rm au}}

\defcitealias{Coleman24FFP}{C24}

\newcounter{qnumber}
\definecolor{forestgreen}{RGB}{34,139,34}

\usepackage{amsmath}
\usepackage{color}

\title[FFP Predictions]{Predicting the Galactic population of free-floating planets from realistic initial conditions}

\author[Coleman and DeRocco]{Gavin A. L. Coleman\thanks{Email: gavin.coleman@qmul.ac.uk}$^{1}$ and 
William DeRocco$^{2,3,4}$
\\
$^{1}$Astronomy Unit, Department of Physics and Astronomy, Queen Mary University of London, Mile End Road, London, E1 4NS, UK\\
$^{2}$Santa Cruz Institute for Particle Physics, University of California, Santa Cruz, CA 95062, USA\\
$^{3}$Maryland Center for Fundamental Physics, University of Maryland, College Park, 4296 Stadium Drive, College Park, MD 20742, USA\\
$^{4}$Department of Physics \& Astronomy, The Johns Hopkins University, 3400 N. Charles Street, Baltimore, MD 21218, USA\\
}

\date{Accepted 2025 January 22; Received 2025 January 14; in original form 2024 July 08}
\pubyear{2025}
\begin{document}
\label{firstpage}
\pagerange{\pageref{firstpage}--\pageref{lastpage}}
\maketitle
\begin{abstract}
We present the first prediction for the mass distribution function of Galactic free-floating planets (FFPs) that aims to accurately include the relative contributions of multiple formation pathways and stellar populations. We derive our predicted distribution from dedicated simulations of planet birth, growth, migration, and ejection around circumbinary systems and extend these results to also include the contributions from single and wide binary systems. Our resulting FFP mass distribution shows several distinct features, including a strong peak at $\sim8\me$ arising from the transition between pebble and gas accretion regimes and a trough at $\sim 1\me$ due to the shift in the dominant ejection process from planet-planet scattering to ejection through interactions with stars in circumbinary systems. We find that interactions with the central binary in close circumbinary systems are likely the dominant progenitor for FFPs more massive than Earth, leading to a steep power-law dependence in mass that agrees well with existing observations. In contrast, we find planet-planet scattering events in single and wide binary systems likely produce the majority of planets at Mars mass and below, resulting in a shallower power-law dependence. Our results suggest that existing extrapolations into the sub-terrestrial mass range may significantly overestimate the true FFP abundance. The features we predict in the mass distribution of FFPs will be detectable by upcoming space-based microlensing surveys and, if observed, will provide key insight into the origins of FFPs and the environments in which they form.
\end{abstract}


\section{Introduction}
\label{sec:intro}

Free-floating planets (FFPs), i.e. those that are not gravitationally-bound to a star, constitute one of the most mysterious demographics of exoplanets. Notoriously difficult to detect, it is only in recent years that decades-long observational campaigns have begun to uncover hints of this population \citep{Mroz2017,Mroz2018,Mroz2019,Mroz2020a,Mroz2020b,Ryu2021,Kim2021,Jung2024}. These few initial detections point to a large abundance of these ``rogue worlds,'' with current estimates suggesting that at Earth mass and below, free-floating planets may dramatically outnumber their bound counterparts \citep{Sumi23}, constituting the majority of exoplanets in this mass range.

However, despite these tantalizing first clues, little is known about the characteristics of this population. Many different processes have been proposed that may contribute to the formation of FFPs, each of which would leave an imprint on the mass, velocity, and spatial distribution of the resulting population. Formation mechanisms generically fall into two categories: (1) ``star-like'' and (2) ``planet-like.'' The former refers to processes by which the FFP forms in situ, unbound from any stars at its birth, e.g. by direct collapse or aborted gas accretion onto a stellar core \citep{MiretRoig22}. 
The latter category, ``planet-like'' formation, refers to processes by which an initially bound planet is liberated from its natal star system. Simulations have shown that gravitational scattering events during the chaotic early phases of system formation efficiently eject planets from their birth systems, resulting in a broad and abundant distribution of free-floating planets. Multiple processes have been shown to contribute to this population, including planet--planet scattering \citep{Chambers1996,RasioFord1996,Weidenschilling96,Veras12,Ma16}, stellar flybys \citep{Wang24}, and interactions with binary stars in circumbinary systems \citep{Nelson03,Smullen16,Sutherland16,Coleman23,Coleman24,Standing23,Chen24,Coleman24FFP}. The resulting distribution of FFPs therefore provides a unique snapshot into an early period of system evolution that is otherwise observationally challenging to explore.

With the launch of upcoming space-based microlensing surveys such as the Nancy Grace Roman Space Telescope \citep{Spergel15,Bennett18} and Earth 2.0 \citep{Ge2022}, the FFP mass function will be measured across a wide range of masses \citep{Johnson2020}, providing the first opportunity to explore what this population can reveal about the origins of unbound worlds and their implications on models of planetary formation. To date, little work has been done on connecting specific features in the FFP mass function to corresponding physical parameters associated with their birth. In this paper, we take the first steps to bridge these two. We utilise the simulations presented in \citet{Coleman24FFP} of planetary evolution in circumbinary systems. As described in the following section, these models included the full formation of the objects, including prescriptions for planetesimal/embryo formation \citep{Coleman21}, planetesimal and pebble accretion \citep{Fortier13,Johansen17}, disc-planet interactions \citep{pdk10,pdk11,LinPapaloizou86}, and the accretion of gaseous envelopes \citep{CPN17,Poon21}. From these results, we predict the mass distribution function of FFPs arising from circumbinary systems and extend these results to include single star and wide binary systems as well. We convolve these results with data-driven estimates of the fraction of each system type within the Galaxy to produce a Galactic mass distribution of FFPs. We find that the mass function exhibits key features, such as peaks and troughs, that correspond to differing formation channels. These features, if observed by upcoming surveys, will provide a first glimpse into the physical parameters of the protoplanetary environments that birth FFPs.

This paper is laid out as follows. In Sect. \ref{sec:recap}, we review the principal simulation inputs and salient results from previous studies on which we base the present work. In Sect. \ref{sec:base_model} we outline how we build a galactic population of stars and FFPs based on the results of these simulations. We discuss the resultant combined mass distribution in Sect. \ref{sec:results}, where we combine the populations of FFPs originating in single stars, circumbinary and wide binary systems. In Sect. \ref{sec:comparison} we compare our mass functions to those predicted from observations. Finally in Sect. \ref{sec:conc} we draw our conclusions.

\section{Simulation Dataset}
\label{sec:recap}

We begin by providing a summary of the key results from \citet[hereafter \citetalias{Coleman24FFP}]{Coleman24FFP}, where the authors used a comprehensive planet formation model to simulate where and how planets form in circumbinary discs, placing a specific emphasis on the production of FFPs. The creation of the planet formation model used is outlined in detail in \citet{Coleman23,Coleman24}, but involved attaching a 1D viscously evolving circumbinary disc into the \textsc{mercury6} symplectic N-body integrator \citep{Chambers} updated to accurately model
planetary orbits around a pair of binary stars \citep{ChambersBinary}. The circumbinary discs included prescriptions for a central cavity, carved through dynamical interactions between the gas and the binary stars \citep{Artymowicz94,Dutrey94}, and were allowed to evolve through viscous accretion onto the central stars \citep{Shak,Lynden-BellPringle1974} and by mass loss through photoevaporative winds. The wind model included the high energy radiation originating both from the central stars \citep[internally driven,][]{Ercolano21,Picogna21} and from nearby massive stars in the surrounding environment \citep[externally driven,][]{Haworth23}. Planets were able to form in the circumbinary discs, initially as planetary embryos after the gravitational collapse of drifting pebble clumps \citep{Coleman21}, and then grow by accreting planetesimals \citep{Fortier13}, pebbles \citep{Johansen17} and gas \citep{Poon21} from the surrounding disc. As the planets grew they underwent migration whilst embedded within the circumbinary discs \citep[Type-I,][]{pdk10,pdk11} as well as after having opened a gap in the discs following runaway gas accretion onto their cores \citep[Type-II,][]{LinPapaloizou86}. Collisions between these evolving planets were tracked and treated as perfect inelastic mergers primarily due to the computational expense of including a full fragmentation prescription. However, for the majority of the collisions, the collision velocity was less than 1.5 times that of the mutual escape velocity, which typically results in efficient accretion onto the more massive planet \citep{Agnor04}, hence can be modelled as an effectively perfect merger. The simulations were run for 10 Myr, a choice that captured both the complete evolution of the circumbinary discs, as well as a period of time afterwards during which planets underwent mutual interactions in a gas--free environment.

\citetalias{Coleman24FFP} found that these simulations often eject planets onto unbound orbits, and used the simulation results to explore from what processes such ejections arose. The authors varied the circumbinary separation between 0.05--0.5$\au$ and explored various levels of viscosity. The general conclusion was that in all of the simulations, the formation of the FFPs followed similar pathways: multiple planetary embryos would form in the circumbinary discs, and then grow through the accretion of pebbles, planetesimals and gas. Once these planets reached $\sim$Lunar mass, they would begin to migrate efficiently through the disc, typically inwards towards the central stars. Their migration would stall at the central cavity, where they would continue to grow. Dynamical interactions with other planets would then force some of these planets to interact with one or both of the central binary stars, resulting in their ejection from the system. Additionally, some planets would be ejected from further out in the disc through planet--planet interactions. An example of the temporal evolution of such a system can be found in Figure 1 of \citetalias{Coleman24FFP}. Through these processes, \citetalias{Coleman24FFP} found that the majority of ejections (both planet-star and planet-planet) occurred within the first $\lesssim 3$ Myr of simulation.

The simulation results from \citetalias{Coleman24FFP} provide a rich and physically-motivated dataset from which to study FFPs. As such, in this paper, we use this existing dataset to construct a first model of the overall Galactic population of FFPs, marginalizing over contributions from different ejection processes and stellar architectures, as described in the following section.

\section{Building a galactic distribution of stars and FFPs}
\label{sec:base_model}

\subsection{Early and late-time instabilities}

Planetary ejection can occur throughout the growth and evolution of a planetary system. At early times ($\lesssim 10$ Myr), the rapid growth and migration of bodies during disc dispersal lead to chaotic interactions that result in high rates of collision and ejection. Eventually, the disc disperses and these interactions diminish, resulting in the long-term quasi-stable planetary systems we observe around the majority of stars. However, whilst most instabilities are settled within the first few million years after final disc dispersal, some instabilities take tens of millions of years, or longer, to manifest \citep{Izidoro17}. Such instabilities can arise due to, for example, flybys of nearby stars deep within the gravitational well \citep{Malmberg2011,Wang2020,Wang24}, changes in wide binary orbital parameters due to galactic perturbations \citep{Kaib2013}, or long time-scale secular perturbations between planets \citep{PuWu2015}. Such interactions may ultimately trigger further ejection of planets, contributing to the galactic population of FFPs. It is this late-time regime that the majority of existing numerical studies of FFP formation address. Such studies begin with fully-formed planets placed on orbits drawn either randomly or from some observationally-motivated distribution and follow their evolution over long time-scales \citep[e.g.][]{Juric2008,Veras12,Smullen16}. Often, the initial conditions for such simulations are chosen arbitrarily, hence it is difficult to connect the results to the true expected abundance of FFPs.

Early-time instabilities, on the other hand, have received relatively little attention \citep[e.g.][]{Ma16,Coleman24FFP}, owing in part to the computational overhead associated with simultaneously modeling the growth and evolution of planetary bodies in the protoplanetary disc along with the gravitational interactions that lead to ejection. However, these instabilities likely dominate the FFP abundance, as the majority of dynamical interactions occur on short time-scales, and therefore should occur early in the planet lifetimes \citep{PuWu2015}. See Sect. \ref{sec:conc} for further discussion of early- versus late-time expected yields. As such, in this work, we focus primarily on FFPs formed via early-time instabilities and leave a more detailed study of late-time effects to future work.

\subsection{Choice of models}
\label{sec:dist_prev}

To build our galactic model, we adopt the population of planets from \citetalias{Coleman24FFP} that formed in circumbinary discs with low levels of turbulence (viscous parameter $\alpha=10^{-3.5}$--$10^{-4}$). We use this population since this level of turbulence is consistent with observational estimates \citep{Isella09,Andrews10,Pinte16,Flaherty17,Trapman20,Villenave20,Villenave22}, as well as from constraints placed from theoretical work \citep{Standing23,Coleman24}. Additionally, these models were able to produce planets and planetary systems similar to observed circumbinary systems, e.g. BEBOP-1 \citep{Coleman24}. When comparing the mass distributions of bound objects in these simulations to observed distributions in single star systems \citep{Fernandes19,Hsu19,Zhu21,Ananyeva22}, we also find good agreement. 

\begin{figure}
\centering
\includegraphics[scale=0.55]{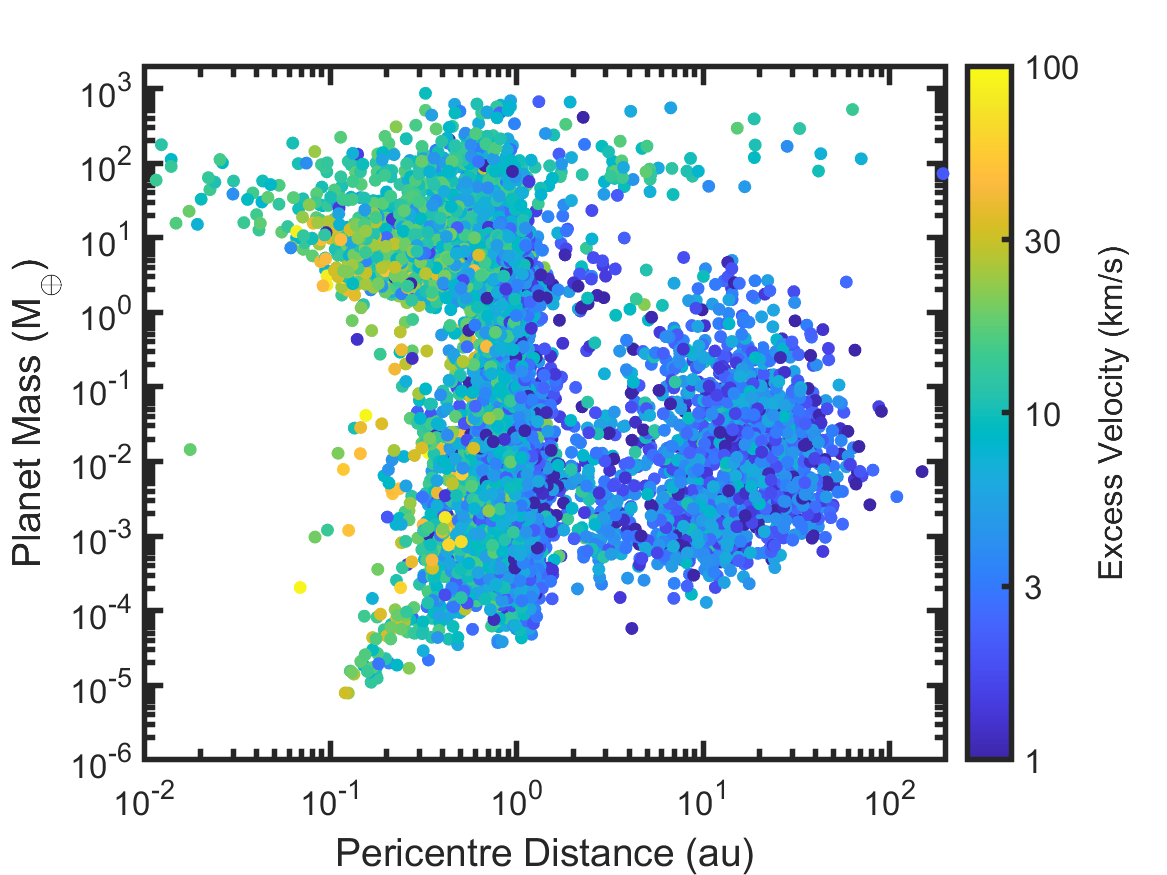}
\caption{Planet mass versus the barycentric pericentre distance for ejected planets for the low $\alpha$ simulations of \citet{Coleman24FFP}. The barycentric pericentre distance gives an approximate location for the final interaction that led to the ejection of the planets. The colour coding shows the excess velocities of the planets after they have been ejected.}
\label{fig:mva}
\end{figure}

Having selected this subset of the \citetalias{Coleman24FFP} models, we can begin to explore the characteristics of the associated FFPs. In Fig. \ref{fig:mva}, we show the masses and ejection locations of these planets. There are two main populations of ejected planets, those close to the star (inner $\sim$few $\au$) that are ejected through interactions with one or both of the binary stars, and those planets ejected far from the central stars through interactions with other planets, with the two populations having different mass distributions.
Additionally the ejection velocities of the planets, indicated by the colour of the points, are also different for the two populations, with planets ejected through interactions with the central binary typically exiting the systems with greater velocities than those that are ejected through planet-planet interactions.
As discussed in \citet{Coleman24FFP}, these greater velocities, averaging around $\sim10\,\rm km/s$, are much larger than the velocity dispersion of stars in star forming regions \citep{Kim19,Theissen23}. As a result, for more massive FFPs, i.e $>1M_{\rm Jup}$, observations of larger velocity dispersions compared to nearby stars could in principle point to planet-like formation origins, rather than star-like formation origins. However, as the planets age,  these differences in velocity dispersions will diminish as gravitational interactions with other objects equalise the velocity dispersions to that seen for stars in different parts of the Milky Way, where the stellar dispersion is on the order of many tens of $\rm km/s$ \citep{Sharma21}.

Though these simulations were performed for circumbinary systems, the planet-planet ejections have very weak dependence on the the system being circumbinary. These planets form at large distances from the central binary and do not undergo significant migration before ejection. At these large orbital distances, the dynamical effects of the binary are negligible, hence the planets evolve and interact as if there were only a single central star. Near to the eccentric cavity of the circumbinary disc, meanwhile, planet-planet scattering tends to be inefficient since the planets are deep into the gravitational well, hence only massive planets are able to eject other objects, whilst interactions between less massive planets would typically lead to collisions. This is supported by multiple previous planet formation works that resolve planet formation processes close to the central star \citep[e.g.][]{ColemanNelson16,ColemanNelson16b,Emsenhuber2021b}. Indeed, when comparing the frequency of planet ejections from those works to this work, we find good similarity, with differences being at most of factor two. This is even considering that the physical models and assumptions are different across these different works. Ultimately, as there seems to be good agreement between single star simulations and the planet--planet ejections here, we assume that the results of the circumbinary simulations presented here serve as a good proxy for the formation and ejection of planets around single star systems.

\begin{table*}
    \centering
    \begin{tabular}{ccccc}
    \hline
        Stellar Population & Semi-major Criteria & Stellar Percentage & FFPs / system & FFP total / star\\
        \hline
        Circumbinary & $a_{\rm b}<3\au$ & 14\% & 9.7 & 1.36\\
        Ultra Wide Binaries & $a_{\rm b}>300\au$ & 15\% & 2.28 & 0.34\\
        Intermediate Binaries & $3\au<a_{\rm b}<300\au$ & 31\% & 0 & 0\\
        Single Stars & - & 40\% & 1.14 & 0.46\\
        \hline
        Total & - & 100\% & 3.12 & 2.16\\
        \hline
    \end{tabular}
    \caption{Contributions from different stellar populations to the Galactic population of FFPs for planets with masses $m_{\rm p}>10^{-2}\me$.}
    \label{tab:pop_ratios}
\end{table*}

\subsection{Incorporating multiple stellar configurations}

As circumbinary systems constitute only a fraction of the Galactic population of star systems, the distribution of ejected FFPs that will arise from the simulations is not fully representative of an actual galactic population of FFPs. To derive a prediction for such a population, we must take into account the relative fraction of binary star systems of varying types and the corresponding ejection rates from single-star systems as well. 

We begin with the binary fraction. 
Observations of binarity for Solar type stars have found that 60\% of stars between 0.75--1.25$\msun$ have a companion \citep{Offner23}, where that companion could be orbiting in a close binary configuration or wide binary configuration. These are two very different dynamical systems, hence estimates of the relative abundance of these systems is also required. Multiple works have explored the distribution of binary stars \citep[e.g.][]{Raghavan10,Kounkel19,Moe19,Winters19}, whilst \citet{Raghavan10} found that the binary separation frequency for FGK stars follows a log-normal distribution, which we employ to estimate the relative abundance of circumbinary systems of varying semimajor axis.

Though the simulations on which we base our results only modelled binary stars with separations 0.05--0.5$\au$, we argue that they are applicable to slightly wider binaries as well. This is due to the fact that the majority of planets that ultimately become FFPs are formed far out in the disc, typically $>10\au$ \citep{Coleman23,Coleman24}. With binary cavities typically extending out to 3--4$\times$ the binary separation \citep{Artymowicz94,Thun17,Coleman22b}, we therefore make the assumption that the resulting distributions of ejected planets would be similar to that presented above for close binary systems of separations $a_{\rm b}\le 3\au$. Extending further than this would result in the cavity beginning to encroach on the sites where planets, particularly planetary embryos, grow most efficiently. This encroachment would then start to hinder planet accretion processes due to larger relative velocities between planets and pebbles \citep{Pierens20,Pierens21}, and larger impact velocities between planets and planetesimals \citep{Paardekooper12,Meschiari12a,Meschiari12b,Lines14,Bromley15}, with both accretion regimes operating at lower efficiencies. With this constraint in mind, and following \citet{Raghavan10} in using a log-normal distribution with $\mu=40\au$ and $\sigma_{\rm log~a}=1.5$ we find that 23$\%$ of binary systems can be classified within our close binary regime. When taking into account the binary fraction of 60$\%$, this results in $14\%$ of all systems being of close binary in nature.

With close binaries and single stars making up 54$\%$ of the stellar population (singles contributing 40\% overall), that leaves wide and intermediate binaries to predict the binary population for. We first look at wide binaries, or in our case ultra wide-binaries, e.g. those with separations $a_{\rm b}>300 \au$.
\footnote{Note that we concentrate here on the effects of binary companions on the formation of planets, rather than the subsequent dynamical evolution of planets interacting with a companion star. It is possible that highly eccentric planets may be ejected by such interactions, however the overall number of such planets is low in the \citetalias{Coleman24FFP} simulations. Though beyond the scope of this paper, in future work we plan to perform dedicated simulations of wide binary systems to accurately determine such dynamical effects.}
For equal-mass binary stars, the truncation radius is approximately a third of their separation \citep{Papaloizou77,Artymowicz94}, which for a 300$\au$ equal mass binary equates to $\sim 100\au$. Looking at Solar mass stars, ongoing research has shown that the effects of a binary companion on forming planetesimals and embryo growth are only significantly felt when the circumstellar discs are truncated to $<100\au$ (Coleman in prep). Disc truncation beyond this threshold begins to deplete the abundance of pebbles necessary for rapid pebble accretion, reducing the growth of protoplanets and limiting their final masses \citep{Zagaria21}. For truncation distances greater than 100$\au$, however, it is found that there are negligible differences in the total mass of planets that are able to form (Coleman in prep). To this end, we assume that for all wide binaries with separations $a_{\rm b}>300\au$, the population of FFPs originating from these systems is equal to that of single stars. Additionally, because there are two stars in these systems, there would also be two circumstellar discs, doubling the number of FFPs originating in the systems as a whole. Again following \citet{Raghavan10} and generating a log-normal distribution for binary separation, we find that 25$\%$ of binaries are in this ultra wide binary regime, equating to 15$\%$ of the stellar population.

While ultra-wide binaries could feasibly be assumed to produce FFPs similarly to single stars, for those binaries of closer separation, it is much more difficult to forecast without performing dedicated simulations. As mentioned above, in wide binary systems with separations $a_{\rm b}<300\au$, planet formation is severely hindered due to the reduced time-scales under which pebble accretion may occur. While this may allow numerous low-mass planets to be ejected from each system through mutual interactions with similar sized objects leading to planet-planet ejections, it is not possible to predict their number frequency from our results here. On the other end of the range of wide binary separations, i.e. those with separations $a_{\rm b} \gtrsim 3\au$, the central circumbinary cavity begins to infringe upon the regions where most of the planetary growth for low-mass objects occur, and so their outcomes are equally unclear. Almost certainly, for binaries with separations between 10--100$\au$, the central cavity would be greater than 40$\au$ on one end, whilst the circumstellar discs will be $\sim30\au$ at maximum for the other end. Within this range, planet formation will be severely hindered, and it would be expected that few, if any, FFPs would be formed and ejected. To this end we assume that intermediate binaries do not produce FFPs, but estimate that if they did, it would be at a much lower frequency that that found for single stars for the stellar population. We will explore whether they are able to produce FFPs, and to what frequency, in future work.

\begin{figure*}
\centering
\includegraphics[scale=0.65]{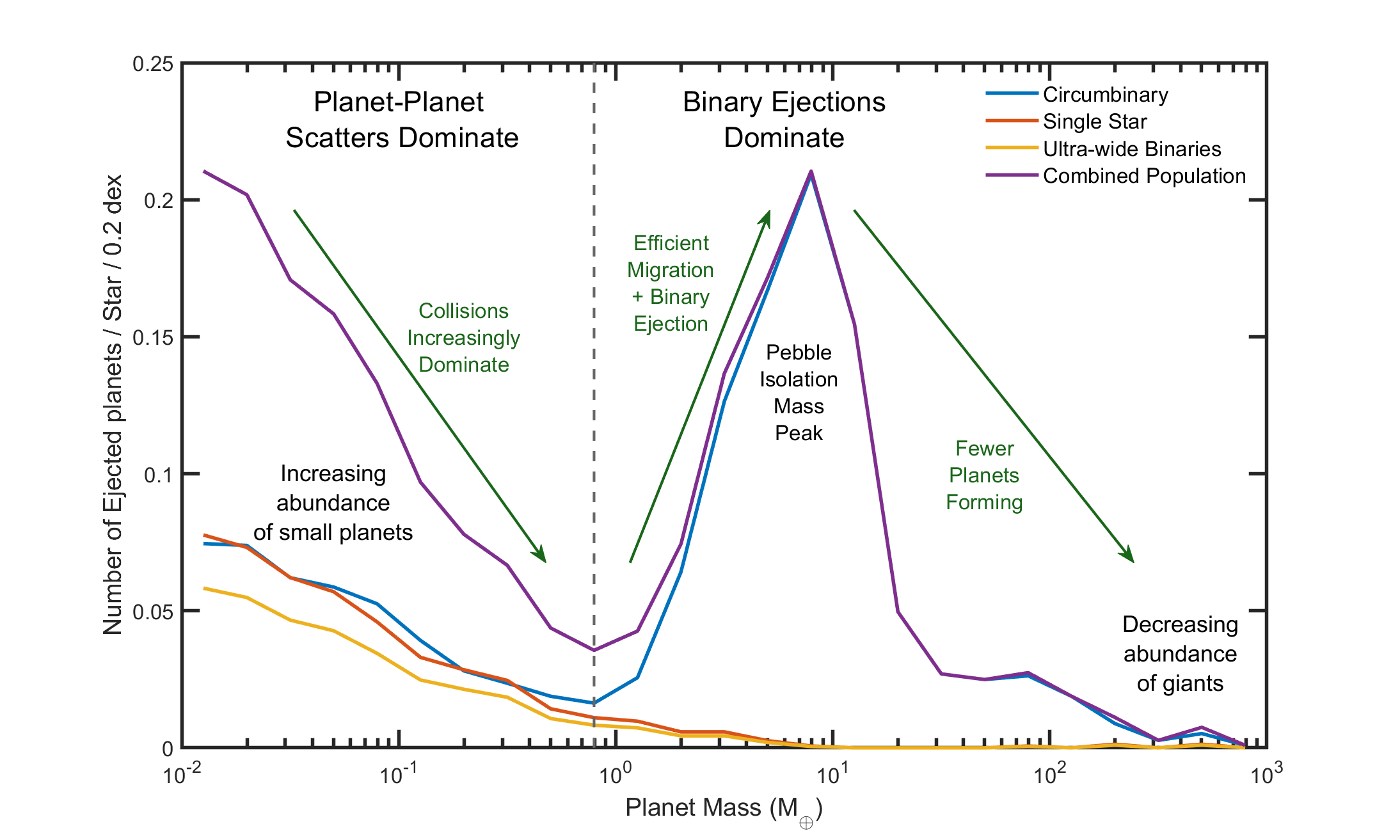}
\caption{The number distribution of ejected planets per star for the galactic population as a function of mass, binned over 0.2 dex. The distributions are shown for planets originating in circumbinary (blue), single star (red), and ultra wide binary ($a_{\rm b}>300\au$, yellow) star systems. A combination of these is shown by the purple line.}
\label{fig:population_distributions}
\end{figure*}

\subsection{Resultant population of FFPs}

With the galactic population of stars described above, we are now in a position to generate the galactic population of FFPs based on our circumbinary simulations.
Figure \ref{fig:population_distributions} shows the resulting mass distribution of FFPs formed per star per 0.2 dex incorporating the relative abundances of various system configurations. Note we do not include intermediate binaries for the reasons discussed in the previous section.
We show the distribution of FFPs ejected by close circumbinary systems (blue line), single star systems (red line), ultra-wide binaries (yellow line), and the whole population (purple line).
In Fig. \ref{fig:population_distributions} we also denote specific features in the distributions with black text, whilst the green arrows and accompanying text show physical trends.
It is clear that the combined distribution contains multiple features and cannot be represented by a single power law function. Notable features include: the peak at super-Earth masses, the lack of terrestrial planets, and the growing number of FFPs as mass decreases into the sub-terrestrial regime.
Below we will discuss the differences between the different populations, before explaining the origins for the specific features in the distributions.

It is clear from Fig. \ref{fig:population_distributions} that circumbinary systems dominate the contribution to the combined population of FFPs. This is especially true for planets more massive than Earth, where circumbinary systems contribute 94.9\% of planets with masses $m_{\rm p}>1\me$. This shows the ineffectiveness of planet-planet scattering for more massive objects. However for lower mass, sub-terrestrial planets, planet-planet scattering becomes more efficient, as these planets are able to scatter off more massive planets. This results in single star and ultra-wide binary systems beginning to contribute significantly to the FFP abundance at low masses. Indeed, for planets with masses $m_{\rm p}< 1\me$, single star and ultra-wide binary systems contribute a combined 62.6\% of those ejected. The relative contribution of single stars and ultra-wide binaries at low masses versus circumbinary systems at higher masses highlights the different primary origins for different mass FFPs, with the transition point between the two different regimes being situated around an Earth mass.

In the latter columns of Table \ref{tab:pop_ratios}, we show the contributions of FFPs per system, and normalised number per star in our galactic population, for the different stellar populations and a totalled contribution in the final row. Cumulatively, by including close binaries, single stars, and ultra wide binaries in our stellar population, we make up $69\%$ of stars. From these stars, they form on average 2.16 FFPs with masses $m_{\rm p}>10^{-2}\me$ per star, with circumbinary systems providing around two-thirds of the distribution.

\section{Discernible features in the FFP distribution}
\label{sec:results}

Here we describe the origins of the features that arise in Fig. \ref{fig:population_distributions}, and discuss the different frequencies of FFPs as a function of planet mass. We will discuss the FFPs arising from the combined population of stars, including circumbinary, single star, and ultra-wide binary systems. This will include both planets that are ejected from the systems by either the binary stars, or through planet-planet scattering.

\subsection{Large abundance of sub-Mars mass objects}
\label{sec:submars}
We begin by looking at planets at the lower end of the mass scale, $m_{\rm p} \le 0.1\me$, corresponding to planets with masses smaller than that of Mars. In this mass range, planet-planet scattering contributes appreciably to FFP formation, and the abundance of FFPs rises smoothly with decreasing mass, with 0.59 planets ejected per star within 0.5 dex of a Mars mass ($m_{\rm p}\sim10^{-1}\me$) and 1.11 planets ejected with masses similar to that of Earth's moon, $m_{\rm p} \sim 10^{-2}\me$. We find that the relative contribution of planet-planet interactions increases uniformly as the planet mass decreases in this range. These interactions take place in the outer disc region, where these low-mass planets are scattered through multiple interactions with growing super-Earth and terrestrial planets, leading to ejection. This is particularly efficient in the outer disc regions since the escape velocity from the more massive planets surface is greater than the escape velocity from the combined stellar system, hence multiple scattering events impart sufficient momenta to eject the smaller body from the system entirely.

The trend of decreasing ejection efficiency with increasing mass in this regime can be explained through a combination of planetesimal abundance and the rate of destructive collisions. Rapid pebble accretion at early times leads to the formation of protoplanetary bodies with masses extending up to $\sim \me$. 
Typically, in planet-planet scattering events that lead to ejection, there is a large mass ratio between the higher-mass body (``ejector'') and lower-mass body (``ejectee''). In order to be successfully ejected, the ejectee must pass close enough to the ejector for the resulting velocity change to exceed the escape velocity of the circumbinary system at the location of the scattering event. For a typical scattering event at $\sim 20 \au$ with ejector mass $\sim \me$ (see Fig. \ref{fig:mva}), the impact parameter within which the ejectee must pass to be ejected is of order $R_\text{ej} \sim 20 \au\times(\me/M_\odot) \sim 1.4 \,R_\oplus$. However, in order to avoid colliding with the ejector, the ejectee must pass at a distance greater than the sum of ejector and ejectee radii. The ejectee radius grows with roughly the cube root of its mass, hence the minimum radius to eject rather than collide decreases with $\approx M_{\text{ejectee}}^{1/3}$. Given that most interactions occur within the plane (i.e. the interacting planets rarely have large relative inclinations), the rate of ejection therefore scales linearly with this minimum radius, and also decreases with $M_{\text{ejectee}}^{1/3}$.

Figure \ref{fig:collisionratio} shows the ratio of collisions to ejections as a function of planet mass, highlighting the growing importance of collisions for low mass planets. For Lunar mass objects, collisions are more prominent than ejections, with around 6 times as many planets being lost through collisions with other objects than are ejected. However as planet mass increases and close encounters become more frequent, the collision to ejection ratio increases. For Mars mass planets, it has already doubled to 12 collisions per ejection. This contributes to the decrease in the number of FFPs as a function of planet mass.
Overall, in the mass range $m_{\rm p} = 10^{-2}$--$10^{-1}\me$, there are a total of 0.87 planets ejected per star.

\subsection{Dearth of Mars to Earth mass objects}
\begin{figure}
\centering
\includegraphics[scale=0.55]{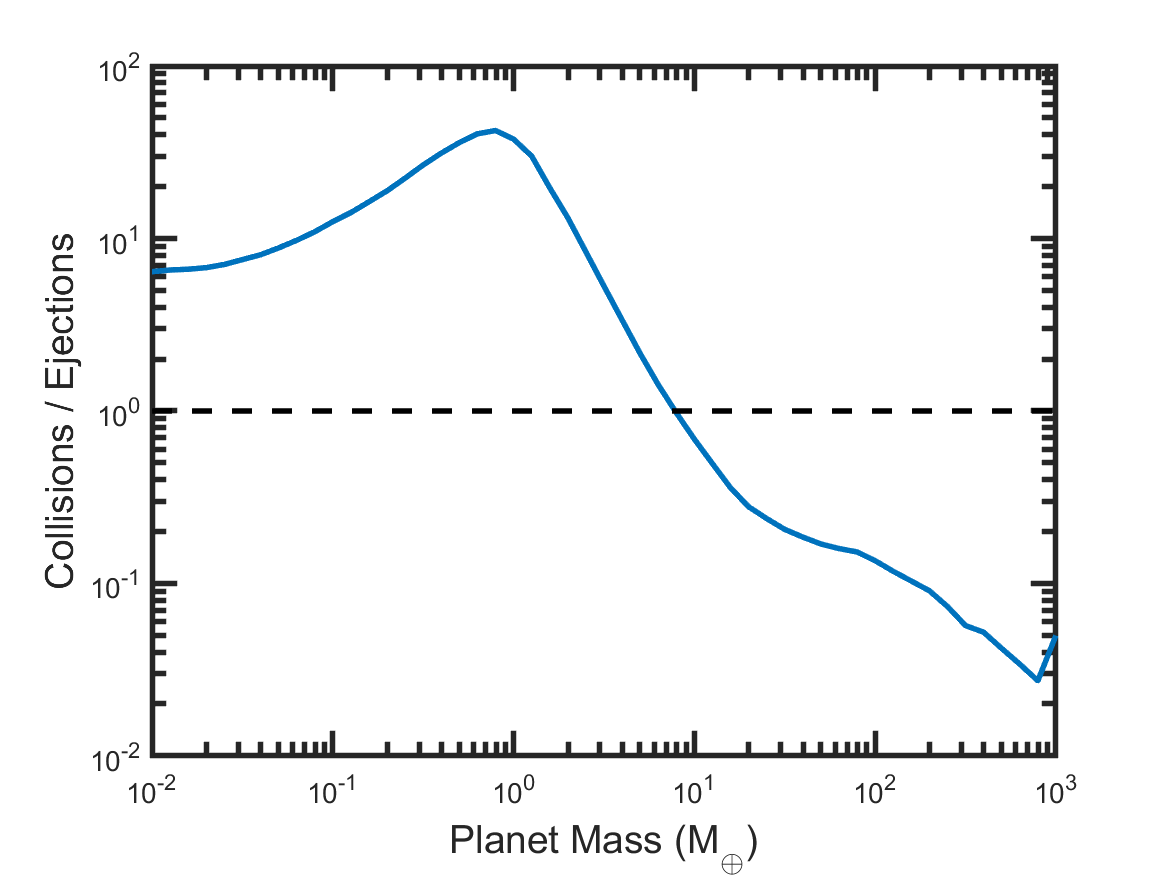}
\caption{Collision to ejection ratio as a function of planet mass. Collisions dominate for lower mass planets, whilst more massive planets above $\sim8\me$ are efficiently ejected during close approaches with other bodies.}
\label{fig:collisionratio}
\end{figure}

Moving further to the right in Fig. \ref{fig:population_distributions},  we find that there is a dearth of planets in the $m_{\rm p} = 10^{-1}$--$10^{0}\me$ mass range. This is due both to the narrow phase space for non-collisional interactions as well as the rapid growth rate of protoplanets at these masses. As can be seen in Fig. \ref{fig:collisionratio},  the peak in the ratio of collisions-to-ejections occurs at $m_{\rm p}=0.7\me$ with 42 planets lost to collisions for every ejection. This dramatically reduces the number of planets able to be ejected around an Earth mass. Furthermore, this mass regime is where pebble accretion is extremely efficient for planets, meaning that if they form early in the disc lifetime, they spend little time in this mass range, reducing their opportunities for interacting with other planets and being ejected.
Interestingly, around an Earth mass ($\pm 0.5\rm~dex$), there are only $\sim0.29$ planets ejected per star, compared to $\sim0.59$ planets with masses similar to Mars. This makes Earth mass planets the \textit{least} abundant demographic of FFPs except for $\approx$ Jupiter-mass gas giants (a mass range that may additionally include a contribution from ``star-like'' formation processes). 

\subsection{Peak at super-Earths}
The next feature that can be seen in Fig. \ref{fig:population_distributions} is the increase in the number of ejected planets at super-Earth masses and the associated peak at $m_{\rm p}\sim8\me$. As planets begin to increase in mass from a Mars to an Earth mass and beyond, the rate of planet migration for embedded planets, i.e. Type-I migration, becomes increasingly efficient, leading to an increasing number of planets being able to migrate towards the circumbinary cavity. The cavity acts as a migration trap whereupon the planets stall, allowing multiple planets to congregate in its vicinity. Subsequent mutual interactions between the planets then force them to interact with one or both of the central stars, often resulting in ejection. Additionally, with pebble isolation masses being around $10\me$ \citep{Ataiee18,Bitsch18}, planets at and above this mass undergo significantly slower growth through gas accretion in comparison to lower-mass planets growing through pebble accretion. As a result, planets tend to remain in this mass range for longer time-scales compared to ejection time-scales, increasing the abundance of FFPs in this mass range.

The fast migration and trapping at the cavity, coupled with the change in accretion regime for these planets leads to the peak in the distribution of ejected planets around $8\me$, as seen in Fig. \ref{fig:population_distributions}, where there are 0.72 planets per star for planets with masses $m_{\rm p} = 8\me \pm 0.5\rm~dex$. When looking at all ejected planets between $10^{0}$--$10^{1}\me$, a total of 0.64 planets are ejected per star. As can be seen in Fig. \ref{fig:population_distributions}, this is by far the most plentiful mass range of ejected planets from planetary systems. Should upcoming space-based microlensing surveys \citep[e.g. Roman,][]{Spergel15,Bennett18} observe this peak in the FFP mass distribution, it would constitute a clear signature of the efficiency of migration traps in circumbinary discs, and additionally of the transition from fast pebble accretion to slow gas accretion, i.e. the pebble isolation mass. With both of those features being dependent on the disc properties, e.g. the turbulence in the disc, a measurement of this peak would provide key insights into the environments in which planets form.

\subsection{Lack of Giant FFPs}
The last main feature of note in Fig. \ref{fig:population_distributions} is the decreasing abundance of ejected planets at masses above that of Neptune. In contrast to the rapid pebble accretion phase, the gas accretion phase proceeds slowly, hence due to the efficient ejection by the binary of planets orbiting near the central cavity, the number of ejected planets that remain bound sufficiently long to reach larger masses is significantly reduced. This reduction in the total number of high-mass planets produces a corresponding reduction in the number of planets that can be ejected from the system, leading to a decreasing abundance of FFPs in the Neptune mass range and beyond. Figure \ref{fig:population_distributions} clearly shows this reduction as the number of planets ejected per binary system drops for planets greater than 10$\me$, with only 0.15 planets ejected between $10^{1}$--$10^{1.2}\me$, down to 0.05 for $10^{1.2}$--$10^{1.4}\me$, and then around 0.026 planets for each bin between $10^{1.4}$--$10^{2}\me$. For even more massive giant planets, the number of planets further decreases, due both to the long time-scales required to grow a planet to these masses as well as the lack of additional objects sufficiently massive to excite a giant planet's eccentricity and drive it inward to be scattered by the binary. Indeed only 0.04 planets above 100$\me$ are ejected per star system on average.

Additionally, we find that as giant planets grow more massive, they once again become efficient at forming FFPs through planet-planet scattering as the cross-section for non-collisional interactions increases with roughly the square root of their mass. Indeed, for planets above 100$\me$, planet-planet scattering accounts for $\sim11\%$ of ejections, whilst for planets between $10$--$100\me$, they only account for $0.4\%$. This shows that though binary interactions still dominate the ejection of FFPs at high masses, planet-planet scattering provides a non-negligible contribution as well, which aligns with the results of previous simulations \citep{RasioFord1996,Moorhead2005,Chatterjee2008}.

As mentioned above, observations of the predicted peak at super-Earth masses and decreasing abundance at higher masses would provide a key indication of the transition out of the pebble accretion regime; furthermore, measurement of the high mass tail would provide an estimate of typical circumbinary disc turbulence. This is because for giant planets to undergo runaway gas accretion, they must remain embedded in the disc for long enough to accrete an envelope equal to or greater than their core mass \citep{CPN17}. This can be interrupted when a planet becomes so massive that it opens a gap in its immediate vicinity in the disc, inhibiting further growth. This gap-opening mass is dependent on the level of turbulence, with more turbulent discs (i.e. larger $\alpha$ viscosities) allowing larger gap-opening masses, hence the formation of more giant planets. Therefore, a measurement of the relative abundance of giant-mass FFPs to Earth-mass FFPs would provide a novel probe of the gap-opening mass and turbulence properties of protoplanetary discs.

\section{Comparison to observed mass functions}
\label{sec:comparison}

Whilst the section above discussed the different features that can be found in our distribution of FFPs, it is also interesting to compare our predicted distributions to existing observations. The majority of FFPs are too dim to be imaged directly, hence are only detectable through their gravitational interactions via microlensing \citep{Paczynski86}. Several different ground-based microlensing surveys have searched for FFPs and have yielded a small number of FFP candidates with masses extending down to $\approx 0.1\,\me$ (see \citet{Mroz2024} for a review).
Such observational studies have attempted to place constraints on the number of FFPs per star \citep{Sumi11,Mroz2017,Gould2022}, suggesting a potential high abundance of low-mass FFPs.

\begin{figure}
\centering
\includegraphics[scale=0.55]{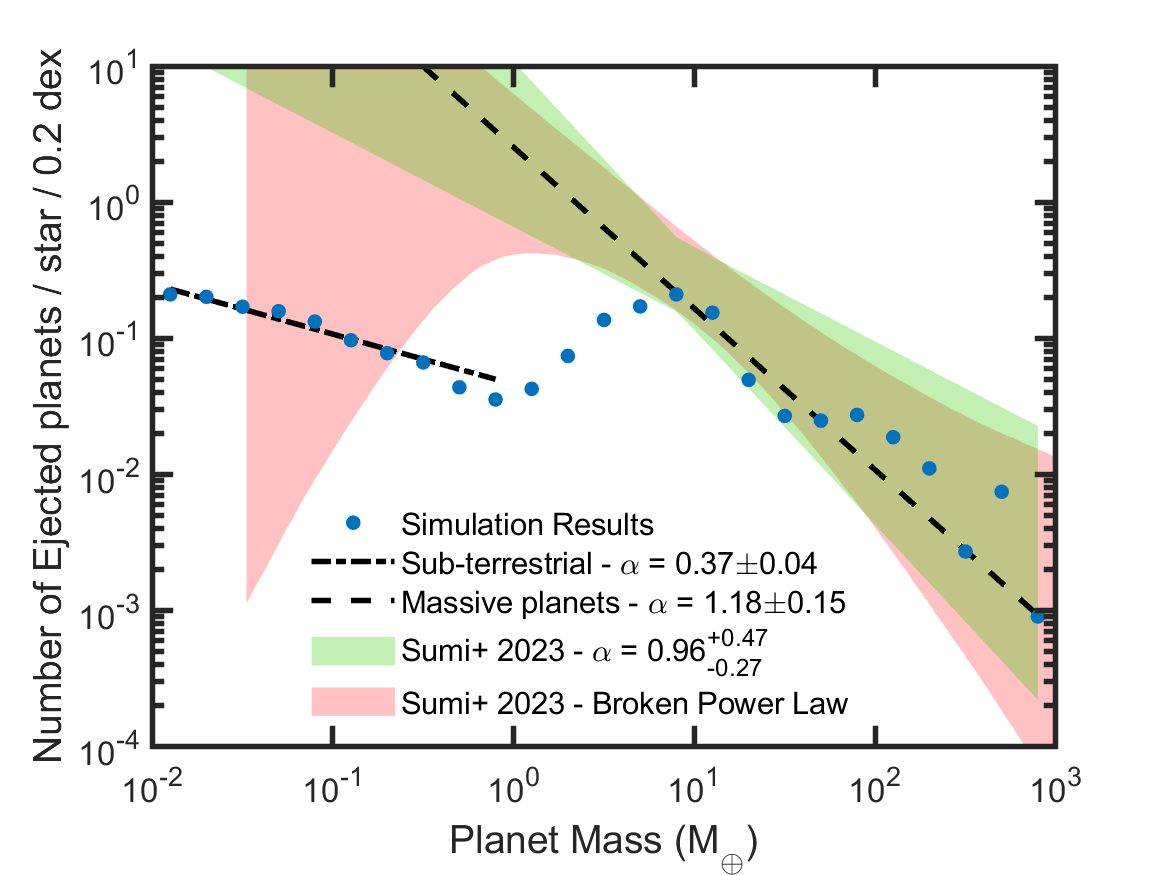}
\caption{Number of ejected planets per star as a function of mass, binned over 0.2 dex. The blue points show the combined distribution arising from our simulations (purple line in Fig. \ref{fig:population_distributions}). The shaded regions show the expected distributions from observations \citep{Sumi23} for a broken power law (pink), and a single power law (green). The dashed and dotted-dashed lines show the power-law fits to our distribution for planets above 8$\me$, and planets below 1$\me$ respectively. All distributions follow a power law $dN/d\log m_p \propto m_p^{-\alpha}$, where the values for $\alpha$ are shown in the legend. See Table \ref{tab:power_laws} for the other best-fit values.}
\label{fig:comparison}
\end{figure}

Recently, \citet{Sumi23} used the MOA microlensing survey toward the Galactic bulge in the 2006--2014 seasons \citep{Koshimoto23} to estimate that for every star in the Galaxy there would be $22^{+23}_{-13}$ FFPs between the masses of $0.33\me<m_{\rm p}<6600\me$, significantly higher than our prediction for the same mass range (1.07 planets per star). Additionally, they predict that the mass distribution of FFPs can be well modelled with a power-law mass function,
\begin{equation}
\label{eq:power_law}
    \frac{dN}{d\log m_{\rm p}}=Z\times\left(\frac{m_{\rm p}}{8\me}\right)^{-\alpha}\rm~dex^{-1}~star^{-1}, 
\end{equation}
where $Z=2.18^{+0.52}_{-1.40}$ and $\alpha=0.96^{+0.47}_{-0.27}$. Note that \citet{Sumi23} find similar results when using a broken power-law, however their uncertainties are significantly increased, hence the resulting fit is primarily useful for displaying more conservative uncertainties on the mass function \citep{KoshimotoPersonal}. 
In Fig. \ref{fig:comparison}, we plot the $1\sigma$ uncertainty from both the \citet{Sumi23} single power-law fit (green shading) with the conservative broken power-law uncertainties (pink shading) alongside our predictions (blue dots). 

Whilst we showed above that our models only predict $\sim2.16$ planets per star (1.07 planets with masses $0.33\me<m_{\rm p}<6600\me$), much less than predicted by \citet{Sumi23}, this is due primarily to the single power-law model adopted by \citet{Sumi23}. They fit their power-law to existing observations consisting of planets $\gtrsim 1\me$, and extrapolate this dependence towards lower masses. Our predictions, which show a dramatic decrease in the FFP abundance below the super-Earth peak, suggest that such an extrapolation likely significantly overestimates the abundance of low-mass FFPs. However, while our predictions disagree in the sub-terrestrial regime, there is interestingly good agreement in the $\gtrsim 8M_\oplus$ regime. In this regime, our models predict 0.53 planets per star, whereas \citet{Sumi23} predict 1.23 planets per star, highlighting the agreement between our models and observations for more massive planets.

\begin{table}
    \centering
    \begin{tabular}{ccc}
    \hline
        Population & Z & $\alpha_4$\\
        \hline
        \citet{Sumi23} & $2.18^{+0.52}_{-1.40}$ & $0.96^{+0.47}_{-0.27}$ \\
        \hline
        $m_{\rm p} > 8\me$ & $1.06\pm0.07$ & $1.18\pm0.15$ \\
        $1\me < m_{\rm p} < 8\me$ & $1.11\pm0.09$ & $-0.72\pm0.11$ \\
        $m_{\rm p} < 1\me$ & $0.11\pm0.02$ & $0.37\pm{0.04}$ \\
        \hline
    \end{tabular}
    \caption{Derived parameters for power law fits to our model, as well as that of \citet{Sumi23}. Our predictions are consistent with existing observations in the high-mass regime, but predict significantly fewer FFP in the unexplored low-mass regime.}
    \label{tab:power_laws}
\end{table}

From Fig. \ref{fig:population_distributions}, it is clear that our combined distribution can be modelled with multiple power-law functions akin to eq. \ref{eq:power_law}, with three distinct populations: low mass ($m_{\rm p}<1\me$), intermediate mass ($1\me < m_{\rm p} < 8\me$), and high mass ($m_{\rm p} > 8\me$).
Note that these three populations arise from three distinct physical regimes: (1) planet-planet interactions with increasing collisional cross-sections, (2) migration of the fast-growing planets towards the central cavity in the discs, and (3) ejection of planets around the cavity during the gas accretion phase. Table \ref{tab:power_laws} shows the derived values for $Z$ and $\alpha$, as well as those values found in \citet{Sumi23}. With the values derived in \citet{Sumi23} being most appropriate for higher mass objects, where the majority of existing observations lie, it is most relevant to compare those values to the high mass population from our models, i.e. $m_{\rm p} > 8\me$. Whilst the normalisation factor $Z$ that we predict has a slightly smaller value than that found in \citet{Sumi23}, it is still within $1\sigma$. More interestingly, there appears to be reasonable agreement on values of $\alpha$, the slope of the power law. This provides an indication that the population observed by existing microlensing surveys may correspond primarily to FFPs ejected from circumbinary systems via planet-binary scattering; this prediction would be confirmed if upcoming microlensing surveys detect the peak and trough structure we predict in the terrestrial-mass range.

In Figure \ref{fig:comparison}, we plot our best-fit power-law models for the high-mass (dashed) and low-mass (dot-dashed) FFP populations. The agreement between our fit and the \citet{Sumi23} power-law is clear for the higher mass objects, whilst the change in origins for FFPs accounts for the differences in the low mass population. Indeed, for lower mass objects, the slope is significantly shallower than that predicted from observations and aligns with the expectation of a roughly $m_p^{-1/3}$ dependence (Sect. \ref{sec:submars}). The number of Earth-mass FFPs from simulations is roughly an order of magnitude lower than the prediction from observations, further compounding the large difference for low-mass objects. Should future observations show evidence for this peak, its location and relative scale would significantly inform our understanding of planet formation processes and the properties of the disc in which these processes occur.

\section{Discussion and Conclusions}
\label{sec:conc}

In this work we have explored the mass distribution of FFPs arising from a realistic galactic population of stars. We have used the results from the simulations presented in \citet{Coleman24FFP} that showed that FFPs could easily form in planet formation scenarios within circumbinary systems and have expanded on them by calculating the mass distribution function of planets from a galactic population including predictions for single star, circumbinary, and wide binary systems. We then compare our distributions with those observed from microlensing surveys. The main results from our study can be summarised as follows:

(1) The mass distribution of FFPs originating in circumbinary systems exhibits multiple features including a peak in the distribution at $\sim8\me$, a trough at terrestrial masses, and a power-law dependence at sub-terrestrial masses. The different features correspond to different physical processes affecting planetary ejection in each of those mass ranges.

(2) A peak in the distribution at $m_{\rm p}\sim8\me$. This peak arises after planets have grown to the pebble isolation mass, and migrated in towards the central cavity in circumbinary discs. The planets' subsequent slow growth through gas accretion gives them sufficient time to interact with the central binary stars and be ejected from the system. Observations of this peak would give clear indicators of both the pebble isolation mass, as well as stalling of planet migration at the outer edge of the central cavity.

(3) A dearth of planets ejected in the Earth-mass range. This is due to the dominance of collisions over ejections during close encounters for bodies in this mass range, as the increased planet radii significantly diminish the phase space available to undergo non-collisional interactions. As planet mass decreases, so does the collision-to-ejection ratio. Lower-mass planets are increasingly likely to attain large changes of velocity through multiple interactions culminating in ejection without undergoing a collision.

(4) For giant planets, our populations predict that there are 0.04 FFPs per star, which is well below the observed upper limit of 0.25 \citep{Mroz2017}.  However the true abundance of super-Jupiter FFPs may also receive contributions from star-like formation processes, the modelling of which is beyond the scope of this work.

(5) Looking at the full predicted FFP population as a whole, we find that there is on average 1.07 FFPs per star in the Galaxy with masses, $0.33\me<m_{\rm p}<6600\me$. This prediction is significantly smaller than preliminary estimates that have been made from observations, \citep[e.g. 22 per star,][]{Sumi23}. However, we find that the power-law mass function adopted in those estimates, while constituting a good fit to the high-mass FFP population, likely dramatically overestimates the low-mass FFP abundance.

(6) We find that our number distribution of FFPs follows two power laws, one for high mass planets above $8\me$ and one for low mass planets below $1\me$, with a non-monotonic peak and trough structure connecting the two. The power law for the super-Earths and higher mass planets agrees within present uncertainties with that of \citet{Sumi23}, while the slope we predict for the lower mass planets is much shallower. Observations of these slopes, as well as the peak and trough between them, will give significant information about the origins of FFPs of different masses, as well as the processes that form them.

Ultimately, our models provide the first predictions for the frequency of FFPs, along with their resultant mass distributions, that are derived from realistic initial conditions and take into account a galactic distribution of stellar system architectures. Our results are therefore directly comparable to the mass distribution that existing and future microlensing surveys aim to reconstruct.
Current estimates are only based on a handful of planet candidates from ground-based microlensing surveys \citep{Gould2022,Sumi23}, and so it is difficult at present to effectively compare observed and theoretical distributions of FFPs. However there is good agreement on some features of the distributions, e.g. the slope of the mass distribution for high-mass objects. Future observations with space based missions such as the Roman Space Telescope \citep{Spergel15,Bennett18} and Earth 2.0 \citep{Ge2022} will determine how well theoretical simulations are able to match reality. Additionally, they should be able to verify the features we predict in the distributions, such as the peak in the super-Earth regime, and the trough at terrestrial masses. Determinations of these features, if only qualitatively, will validate the formation pathways of FFPs, whilst quantitative differences will give valuable insights into planet formation processes.

Whilst this work has mainly focused on the formation of FFPs in the first 10 Myr of a systems life, we have not explored the production of FFPs over long time-scales. As systems age, further planet--planet scattering can increase the number of FFPs. To test whether such effects could lead to an appreciable change in our predictions, we allowed a small sample of the \citetalias{Coleman24FFP} simulations to run for an additional 10 Myr (overall 20 Myr), and found that the total number of planets ejected increased by 4\%. However, we found that the rate at which FFPs were being produced decreased over time, since the number of remaining bound planets was decreasing, reducing the frequency of dynamical interactions in the systems that could lead to ejections. Additionally, we found the mass distributions of the ejected planets at later times to be broadly consistent with those that were ejected at early times. As such, whilst we expect the number of FFPs to increase marginally over time, we do not expect such effects to lead to large qualitative changes to the distributions shown in Fig. \ref{fig:population_distributions}.

In addition to the long term dynamical evolution between planets that can increase the number of FFPs, interactions over short and long time-scales with outer companions in a system, i.e. a wide binary system, could also increase the production rate of FFPs. In this work, we made the assumption, based on planetesimal production and pebble accretion rates, that circular equal mass binary systems with separations greater than 300 $\au$ did not affect the planet formation process. This neglected the dynamical interactions between an outer companion and orbiting planets, especially those with longer periods. Such dynamical interactions with the binary would most likely eject such planets early in the life of the system, within the first 100 Myr. In future work, we will explore the effects of an outer companion on the dynamical evolution of planets in the outer systems, and on the production rates of FFPs. It is worth noting, however, that even if all of the planets on long-period orbits in our simulations were ultimately ejected, this would only add on average a single planet per star in each wide binary system, only slightly increasing the number of FFPs overall. Interactions with other stars in stellar clusters, i.e. those stars performing flybys, can also increase the rate of FFP production over long time-scales, however these effects are once again expected to only appreciably affect the long-period planets discussed above. A more quantitative study of these effects will be explored in future work.

The simulations that provided the population of FFPs here were based on the results of planet formation simulations for a circumbinary system with a specific binary central mass, mass ratio, and eccentricity. Those simulations did not include different stellar masses or binary orbital properties, and so there remain several opportunities to improve our model with a more complete sample of system architectures in the Galaxy. In future work, we will perform simulations of a full stellar population of single and binary stars, exploring the outcomes of different stellar masses, binary orbital parameters, and multiplicity as a function of mass \citep{Duchene13}, as well as varying initial conditions and star-forming environments for their respective protoplanetary discs. This will allow us to refine our model of the FFP population as a whole, whilst also being able to explore how the population changes with different stellar populations. Already however, our results make clear predictions for the FFP mass distribution that will be tested by future microlensing surveys; if observed, such structures will provide a new window into the origins of planets and their ultimate fates.

\section*{Data Availability}
The data underlying this article will be shared on reasonable request to the corresponding author.

\section*{Acknowledgements}
The authors thank the anonymous referee and Alex Mustill for providing useful and interesting comments that improved the paper.
GALC acknowledges funding from the Royal Society under the Dorothy Hodgkin Fellowship of T. J. Haworth, STFC through grant ST/P000592/1, and the Leverhulme Trust through grant RPG-2018-418.
WD acknowledges the support of DOE grant No. DE-SC0010107.
This research utilised Queen Mary's Apocrita HPC facility, supported by QMUL Research-IT (http://doi.org/10.5281/zenodo.438045).
This work was performed using the Cambridge Service for Data Driven Discovery (CSD3), part of which is operated by the University of Cambridge Research Computing on behalf of the STFC DiRAC HPC Facility (www.dirac.ac.uk). The DiRAC component of CSD3 was funded by BEIS capital funding via STFC capital grants ST/P002307/1 and ST/R002452/1 and STFC operations grant ST/R00689X/1. DiRAC is part of the National e-Infrastructure.
The authors would like to acknowledge the support provided by the GridPP Collaboration, in particular from the Queen Mary University of London Tier two centre.

\bibliographystyle{mnras}
\bibliography{references}{}

\label{lastpage}
\end{document}